\documentclass[journal]{IEEEtran}

\usepackage{cite}
\usepackage{amsmath,amssymb,amsfonts}
\usepackage{algorithmic}
\usepackage{graphicx}
\usepackage{textcomp}
\usepackage{hyperref}

\def\BibTeX{{\rm B\kern-.05em{\sc i\kern-.025em b}\kern-.08em
T\kern-.1667em\lower.7ex\hbox{E}\kern-.125emX}}

\begin{document}

\title{GHOST-CAT: An Efficient and Practical Network for Mesh Generation from 3D Echocardiography}

\author{
\IEEEauthorblockN{
Edward Ferdian\IEEEauthorrefmark{1}\IEEEauthorrefmark{2},
Debbie Zhao\IEEEauthorrefmark{1},
Alistair A. Young\IEEEauthorrefmark{3},
and Martyn P. Nash\IEEEauthorrefmark{1}\IEEEauthorrefmark{4}
}

\vspace{1em}

\small
\IEEEauthorblockA{\IEEEauthorrefmark{1}
Auckland Bioengineering Institute, University of Auckland, Auckland, New Zealand}
\IEEEauthorblockA{\IEEEauthorrefmark{2}
Faculty of Informatics, Telkom University, Bandung, Indonesia}
\IEEEauthorblockA{\IEEEauthorrefmark{3}
School of Biomedical Engineering and Imaging Sciences, King's College London, London, United Kingdom}
\IEEEauthorblockA{\IEEEauthorrefmark{4}
Department of Engineering Science and Biomedical Engineering, University of Auckland, Auckland, New Zealand}

\IEEEauthorblockA{
\textit{Corresponding emails:} efer502@aucklanduni.ac.nz, debbie.zhao@auckland.ac.nz}

\vspace{1.5em}
\normalsize
\IEEEauthorblockA{
\textit{Preprint submitted to IEEE Transactions on Medical Imaging}}

\thanks{This study was funded by the Health Research Council of New Zealand (programme grant 23/527). D. Z. is supported by a Pūtahi Manawa \(|\) Healthy Hearts for Aotearoa New Zealand Centre of Research Excellence Research Fellowship (Grant Number FLW23-001).}
}

\maketitle

\begin{abstract}
Recent advances in deep learning have significantly accelerated cardiac imaging workflows, from segmentation to the generation of meshes for computational modelling. Nevertheless, analysis of 3D echocardiograms presents unique challenges due to their low contrast-to-noise ratio, conical field of view, and susceptibility to acoustic shadowing. Here, we present an efficient and practical network tailored for 3D echocardiograms. Our method consists of a two-stage network that combines convolutional neural networks, graph convolutional networks, and transformers, to create accurate time-varying 3D meshes of the left ventricle that are topologically consistent and temporally coherent throughout the cardiac cycle. Our model achieved superior mesh reconstruction accuracy compared to current state-of-the-art methods on a held-out test dataset of 100 3D echo images, with a Dice coefficient of 0.87~±~0.05 (cavity) and 0.75~±~0.07 (myocardium), and mean~±~SD surface distances of 3.3~±~0.6~mm (endocardium) and 3.5~±~0.5~mm (epicardium), against reference segmentations derived from cardiac magnetic resonance imaging. The reconstructed mesh enables automated calculation of routine clinical indices, such as volume, mass, and strain, and enables advanced applications with biophysical digital twins. Source code is openly shared at 
\url{https://github.com/EdwardFerdian/ghost-cat}.
\end{abstract}


\section{Introduction}
As the most widely used cardiac imaging modality, echocardiography (echo) provides real-time imaging of the heart in a non-invasive and efficient manner. Technological advancements in 3D echo imaging have enabled direct volumetric assessment of the heart without geometric assumptions present in conventional 2D imaging. However, the analysis of 3D echo can be particularly difficult due to the poor contrast-to-noise ratio, a conical field of view (which can result in parts of the heart being cropped), and susceptibility to acoustic dropout. As a result, manual analysis of 3D echo is highly variable and reader dependent. To overcome these challenges, several deep learning solutions have been proposed. Convolutional neural networks (CNNs) are commonly used to extract rich features from images, with nnU-Net~\cite{Isensee2021NnU-Net:Segmentation} being a leading example. Recent foundation models, such as the Segment Anything Model (SAM)~\cite{kirillov2023segany}, have also been implemented for medical images~\cite{MedSAM}, and fine-tuned for various modalities and anatomical structures~\cite{cheng2023sammed2d, wang2023sammed3d, Lei2023medlam}. While such techniques have demonstrated marked improvements in automated analysis, accurate and reliable segmentation of 3D echo images remains a challenging area of research~\cite{Zhao2023MITEA:Imaging, 10306804, 10054471}.

The segmentation of structures of interest alone enables the quantification of routine clinical metrics such as chamber volume and myocardial mass. However, meshes with consistent topologies are required for more advanced research and clinical applications, such as modelling cardiac electrical activation and biomechanics~\cite{Wang2018LeftAnalysis}, statistical shape analysis~\cite{Gilbert2019IndependentStudy}, and the derivation of kinematic indices such as strain~\cite{GORCSAN20111401}. Furthermore, temporal coherence is needed to enable reliable characterisation of systolic and diastolic motion patterns. In addition to routine clinical indices, insights and indices derived from advanced modelling techniques may provide added value for diagnosis and prognosis. Recent advances have focused on mesh reconstruction by combining CNNs for image feature extraction with with graph convolutional networks (GCNs)~\cite{Kipf2017Semi-supervisedNetworks} that are able to operate directly on mesh structures. Although these methods are generally robust for high-resolution modalities such as computed tomography and cardiac magnetic resonance imaging, no existing method specifically addresses the unique challenges posed by 3D echo. To this end, we present GHOST-CAT (Graph-based Hybrid Optimiser for Spatio-Temporal Cardiac Analysis Tasks), a two-stage deep learning network, which produces topologically consistent and temporally coherent meshes of the left ventricle (LV) directly from 3D echocardiograms.

The main contributions of this paper are: 1) a hybrid CNN and GCN approach, inspired by Voxel2Mesh~\cite{10.1007/978-3-030-59719-1_30}, leveraging a topologically consistent template mesh of the LV; 2) a strategy for the generation of pseudo ground truth meshes for transitional image frames (i.e., those not at end-diastole (ED) or end-systole (ES)), for which ground truth meshes are not available; and 3) a second transformer-based network to further process graph features into temporally coherent predictions that promote the conservation of mass of the myocardium across the cardiac cycle.

\section{Related Works}
Conventional workflows for cardiac mesh generation typically adopt a multi-step process, whereby segmentations (in the form of labelled voxels in a dense structured grid) are obtained from images, followed by the generation of polygonal surfaces using marching cube algorithms~\cite{10.1145/37401.37422}. These meshes are subsequently refined during post-processing, followed by landmark identification and geometric fitting to obtain anatomically meaningful surface representations. Such pipelines can be complex, time-consuming, and prone to compounding errors. Consequently, contemporary methods increasingly focus on direct mesh reconstruction from images.

\subsection{Hybrid graph architectures}
Recent work has highlighted the potential of GCNs~\cite{Kipf2017Semi-supervisedNetworks} in image domains. Typically, these are combined with CNNs to first extract spatial features, which are then passed on to the graph networks. Examples include HybridVNet \cite{GAGGION2025103630} and MR-Net \cite{CHEN2021102228}, which leverage hybrid CNN/GCN architectures to generate high-fidelity cardiac meshes from multi-view 2D images. In 3D, notable methods include Voxel2Mesh~\cite{10.1007/978-3-030-59719-1_30}, MeshDeformNet~\cite{KONG2021102222}, and ModusGraph~\cite{10.1007/978-3-031-43990-2_17}. Voxel2Mesh represents one of the pioneering efforts to combine CNNs with GCNs, extending upon the concept of Pixel2Mesh~\cite{DBLP:journals/corr/abs-1804-01654} by adding a voxel decoder that bypasses the need for explicit ground truth meshes. Although reasonably high-quality meshes can be derived using this method, they lack topological consistency, which limits their utility for advanced kinematic analyses and computational modelling. To overcome this, MeshDeformNet~\cite{KONG2021102222} progressively deforms a high-resolution template mesh, imposing point correspondence, while avoiding the need for subdivision steps. This approach, while utilising a different graph convolution variant, maintains the core concept of leveraging GCNs for feature processing from a voxel encoder or decoder, with the added advantage of reconstructing multiple meshes in a single pass. ModusGraph~\cite{10.1007/978-3-031-43990-2_17} further extends this framework by incorporating spatio-temporal graph convolutions to handle time-varying inputs.

\subsection{Motion and temporal coherency}
In addition to anatomically accurate mesh reconstruction, several studies have explored approaches for modelling cardiac motion while providing temporal coherence. Recent developments have exploited attention mechanisms and transformers for applications such as LV motion tracking \cite{AHN2023102711}, surrogate modelling of haemodynamics \cite{10.1007/978-3-031-94559-5_24}, and cardiac point cloud reconstruction \cite{HeartFormer}. In the context of meshes, Mesh4D \cite{10.1007/978-3-032-05325-1_33} introduces a motion-aware transformer-based variational autoencoder that jointly learns shape and motion information from multi-view images, and MeshHeart \cite{Qiao2025} combines GCNs for mesh encoding with a transformer for temporal dynamics to capture the spatio-temporal distribution of cardiac motion.

Despite these advances, existing mesh reconstruction methods have primarily been developed and validated on high-resolution imaging modalities or multi-view 2D datasets. No existing solutions explicitly address the challenges associated with 3D echo, including low contrast-to-noise ratio, limited field of view, and susceptibility to acoustic dropout. Consequently, direct application of existing methods to 3D echo images can result in spatially inaccurate and temporally unstable meshes.

\section{Methods}
 
\subsection{Data preparation}
We leverage MITEA, a publicly accessible 3D echo dataset with subject-specific labels of the LV cavity and myocardium derived from paired cardiac magnetic resonance (CMR) imaging~\cite{Zhao2023MITEA:Imaging}. In addition to the 134 subjects in the original publication comprising 82 healthy participants and 52 patients with acquired cardiac disease, a further 17 subjects (13 with acquired cardiac disease and 4 healthy participants) were incorporated, producing a total of 604 annotated 3D images acquired from 151 participants (with 4 static images per subject: ED and ES frames from 2 independently acquired cardiac cycles). Briefly, time-resolved subject-specific finite element models of the LV were constructed from cine CMR images, from which static geometries could be extracted at ED and ES. Due to differences in temporal resolution and heart rate variability between imaging modalities, rigid registration between the CMR-derived LV geometries and corresponding 3D echo acquisitions could only be reliably carried out at the states corresponding to maximal and minimal cavity volume. Triangulated endocardial and epicardial surface meshes, consisting of 785 vertices per surface sampled from the finite element model (prior to voxelisation), were utilised for the present study. For additional dataset characteristics, the reader is referred to the original MITEA publication~\cite{Zhao2023MITEA:Imaging}.

\subsubsection{Image pre-processing} 
To ensure consistency of input dimensions, the 3D echo images and corresponding labels were interpolated and zero-padded into ${x, y} \in \mathbb{R}^{160\times160\times128}$. Voxel intensities were standardised on a per-image basis by subtracting the mean and dividing by the standard deviation.

\subsubsection{Template mesh generation}
A rotationally symmetric template mesh of the LV myocardium was constructed by combining two half prolate spheroid meshes, with the inner surface (endocardium) scaled to 0.8 of the length and 0.6 of the diameter of the outer surface (epicardium). Each surface contained 28 vertices along each of the circumferential and longitudinal axes, yielding 785 ($28\times28$ + 1 apex point) vertices per surface to match the target topology and joined at the base using triangular faces to form a continuous mesh of the LV myocardium consisting of 1570 vertices. This served as the initial input to be progressively deformed by the network.

\subsection{Network architecture}
GHOST-CAT comprises a two-stage network: 1) a hybrid CNN/GCN segmentation and surface mesh reconstruction network; and 2) a transformer-based network for producing temporally coherent meshes. 

\subsubsection{Stage I segmentation and mesh generation network}
\begin{figure*}[!t]
\centering
\includegraphics[width=0.92\textwidth]{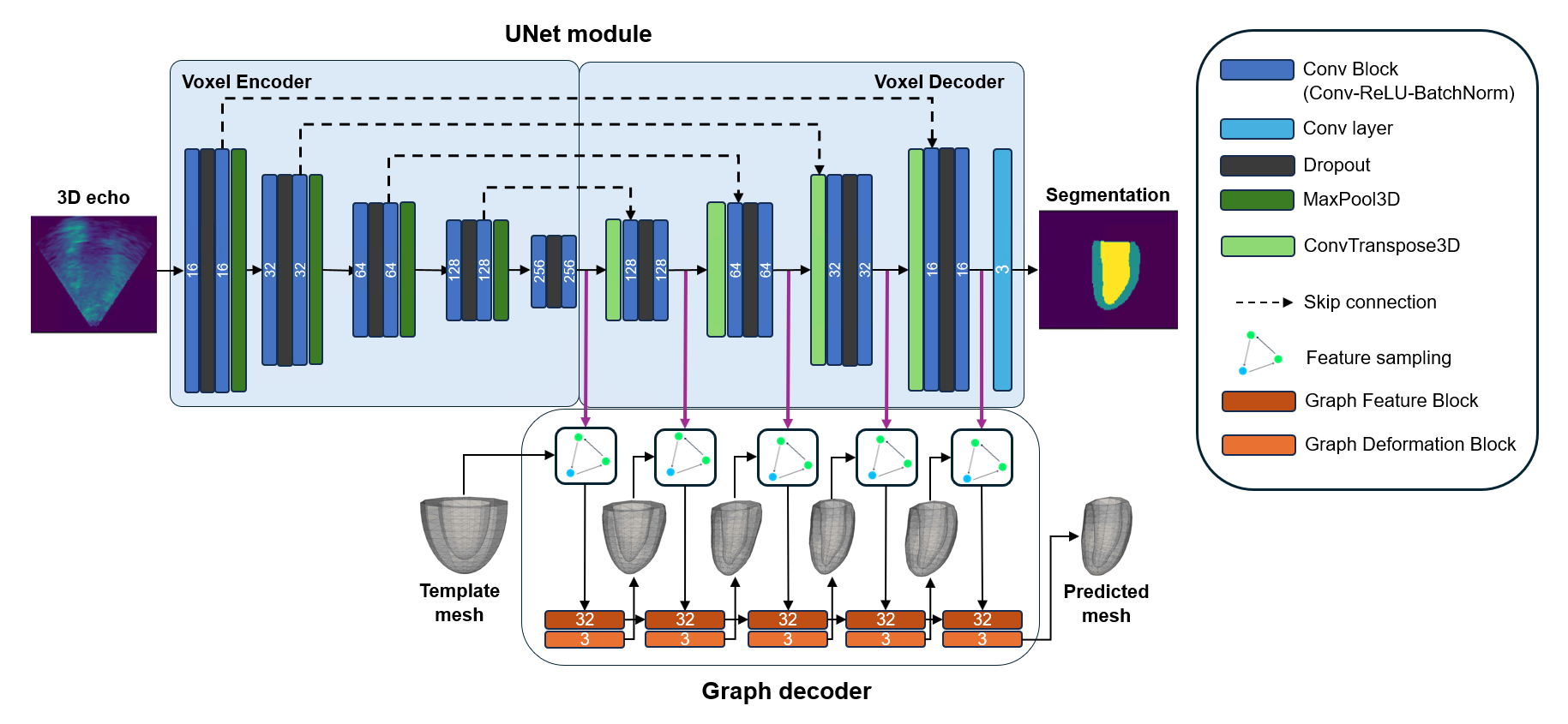}
\caption{GHOST-CAT Stage I network architecture showing UNet module and Graph decoder, which samples features at different resolutions.}
\label{fig1}
\end{figure*}

The Stage I network architecture was inspired by Voxel2Mesh~\cite{10.1007/978-3-030-59719-1_30}, designed to process an image volume and a template mesh to obtain both a voxelised 3D segmentation and surface mesh of the LV myocardium. The Stage I network consists of two main components: a UNet module containing a voxel encoder and decoder, and a graph decoder consisting of feature sampling, Graph Feature Blocks ($g_{conv}$), and Graph Deformation Blocks ($g_{disp}$). A detailed diagram is shown in Fig.~\ref{fig1}.

\textbf{UNet module:}
There are four levels to the UNet encoder, each consisting of two convolution blocks. Each block is made up of a Conv-ReLU-BatchNorm sequence, with dropout layers inserted between blocks, followed by a MaxPooling layer. Similarly, the decoder also contains four levels, with each level consisting of two convolutional blocks applied after upsampling and concatenated with corresponding skip connections from the encoder. This enables the image volume to undergo encoding to latent features, followed by decoding for segmentation reconstruction.

\textbf{Graph decoder:}
The graph decoder deforms the template mesh by utilising features from different resolutions. This process leverages the edges for graph convolutions, following an approach similar to Voxel2Mesh~\cite{10.1007/978-3-030-59719-1_30}. The process begins by sampling feature vectors $x_l$ from voxel features at the coordinates of the vertices belonging to the deformed mesh for each layer \emph{l} (where 1 $\leq$ \emph{l} $\leq$ L). These layers correspond to the voxel decoder blocks. Initially, at layer \emph{l} = 1, voxel features are sampled from the bottleneck layer at the template vertex coordinates. At this stage, $z_0$ represents an empty graph feature vector, and $v_0$ denotes the vertices from the initial template mesh. The Graph Feature Block module then processes the feature vector $x_l$, the previous graph features $z_{l-1}$, and the vertex coordinates $v_{l-1}$ to generate a compact representation of the graph features $z_l$, described by:

\begin{equation}
    zl = g_{conv}(z_{l-1}, x_l, v_{l-1})
\end{equation}

Subsequent layers utilise the graph features $z_l$ to compute the displacements of the vertices through the Graph Deformation Block module. The displacements are then applied to progressively adjust the mesh vertices toward the target mesh at each level, given by:
\begin{equation}
    \bigtriangleup v = g_{disp}(z_{l})
\end{equation}
\begin{equation}
    v_l = v_{l-1} + \bigtriangleup v
\end{equation}

\textbf{Notable modifications from Voxel2Mesh:}
Several important modifications were made to the original Voxel2Mesh architecture to mitigate potential overfitting and maintain feature consistency in preparation for the Stage II network: 

\begin{itemize}
    \item We imposed topological consistency between meshes, therefore necessitating the exclusion of the AdaptiveMeshUnpooling module.
    \item Within the UNet convolution blocks, ReLU activation and batch normalisation were interchanged to ensure zero-centered features were sampled as the input to the graph decoder. Additionally, a dropout layer was introduced between convolution blocks to reduce overfitting.
    \item We utilised a standard feature sampling method with trilinear interpolation at the vertex coordinates. We did not use the LearntNeighbourhoodSampling (LNS), as no improvement in network performance was observed.
    \item In the Graph Feature Blocks, only two graph convolution layers were utilised (instead of the original four) for computational efficiency, with no observed decline in performance.
\end{itemize}

\subsubsection{Stage I loss functions}
The complete loss function for the Stage I network can be expressed as a combination of segmentation loss and mesh loss, as described below.

\textbf{Segmentation loss:}
The segmentation loss function ensures accurate voxel-wise classification, which is crucial to obtain appropriate image features to be sampled by the graph decoder. It combines:
\begin{itemize}
    \item Cross-Entropy Loss ($\mathcal{L}_{ce}$): This term measures how much the predicted probabilities deviate from the true classes.
    \item Weighted Dice Loss ($\mathcal{L}_{dce}$): This term addresses the dissimilarity of the predicted and ground truth segmentation. To help with class imbalance, the myocardium is weighted more than the LV cavity (i.e., myocardium=0.55, cavity=0.45).
\end{itemize}

The total segmentation loss function is then formulated as:
\begin{equation}
    \mathcal{L}_{seg} = \mathcal{L}_{ce} + \mathcal{L}_{dce} 
\end{equation}

\textbf{Mesh loss:}
Unlike the original Voxel2Mesh implementation, where losses are computed at each step of the mesh deformation, our approach computes all losses solely on the final mesh, with the exception of the displacement loss, which is penalised at each level of deformation. The mesh loss function consists of several terms:

\begin{itemize}
    \item Mean squared error (MSE) loss ($\mathcal{L}_{MSE}$): This term measures the difference between the positions of the predicted mesh vertices and the ground truth vertices.
    \item Relative edge loss ($\mathcal{L}_{edge}$): This term promotes consistency between the lengths of associated edges in the predicted and ground truth meshes.
    \item Normal similarity loss ($\mathcal{L}_{norm}$): This term penalises differences in the normal vectors of each of the faces.
    \item Incremental displacement loss ($\mathcal{L}_{disp}$): This term penalises large deformations of mesh vertices between levels, ensuring that the features at each level contribute to the mesh deformation steps.
\end{itemize}

The total mesh loss function is then formulated as:

\begin{equation}
\begin{split}
    \mathcal{L}_{mesh} = {} & \mathcal{L}_{MSE} + \lambda_{edge} \mathcal{L}_{edge} + \lambda_{norm} \mathcal{L}_{norm} \\
    & + \lambda_{disp} \sum_{l=1}^L \mathcal{L}_{disp(l)}
\end{split}
\end{equation}

To balance the contribution of each of the loss terms, the assigned weights were the set to the following: $\lambda_{edge}$ = 0.02, $\lambda_{norm}$ = 0.01, and $\lambda_{disp}$ = 0.05.

\textbf{Overall loss function}
The overall loss function combines both mesh and segmentation losses to balance these aspects during network optimisation using:
\begin{equation}
    \mathcal{L} = \mathcal{L}_{mesh} + \lambda_{seg} \mathcal{L}_{seg}
\end{equation}

where $\lambda_{seg}$ was set to 0.02. This value was determined empirically to balance the contribution of each term for optimal model performance. 

\subsubsection{Stage II temporal network}
\begin{figure}[!t]
\centerline{\includegraphics[width=0.46\linewidth]{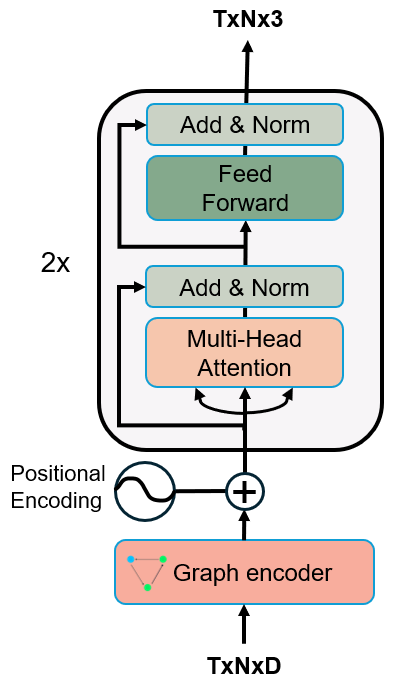}}
\caption{GHOST-CAT Stage II network architecture.}
\label{fig2}
\end{figure}

The Stage II network (Fig.~\ref{fig2}) adopts a transformer architecture~\cite{Vaswani2017AttentionNeed}, which receives the graph features generated by the Stage I network as input. These features are represented as $x_{feat} \in \mathbb{R}^{T \times N \times D}$, where $T$, $N$, and $D$ correspond to the number of time frames, vertices, and feature dimensions, respectively. The maximum number of time frames was set to be 64. Where the input consisted of fewer than 64 frames, a padding mask was applied to disregard computation on missing frames. Conversely, where the maximum number of frames was exceeded, the sequence was truncated by selecting a random index (chosen uniformly between the first frame and the $(T - 64)^{\text{th}}$ frame) and retaining the subsequent 64 frames. As a result, the ES frame was generally included within the sequence, while ED was not necessarily captured. Given the high dimensionality of features, $D$, a shared graph encoder with two graph convolution layers with ReLU activation was used to reduce the dimension to an embedding $x_{emb} \in \mathbb{R}^{TxNx4}$. This embedding, along with positional encoding, was then fed into a transformer with two encoder layers to generate a temporally coherent surface mesh. The resulting output $y \in \mathbb{R}^{TxNx3}$, corresponds to the mesh coordinates at each time frame.

\subsubsection{Stage II loss functions}
The Stage II network was optimised using a combination of MSE and cavity volume, with an added regularisation term to promote a consistent myocardial volume (representing tissue mass) across all frames. To prevent the myocardial volume from collapsing to zero due to mesh intersections between the inner and outer surfaces, the myocardial volume from the first frame of the ground truth sequence was used as the reference. % Note that while the first frame generally corresponds to ED, this is not necessarily the case when the truncation augmentation is activated. 
The resulting loss function is defined as follows:

\begin{equation}
\begin{split}
    \mathcal{L} = \frac{1}{T} \sum_{t=1}^T \Big( \mathcal{L}_{mse} & + \lambda_{cav} \mathcal{L}_{cav} \\ 
    & + \lambda_{myo} (Vol_{myo(t)} - Vol_{myo(0)}) \Big)
\end{split}
\end{equation}

with $\lambda_{cav}$ and $\lambda_{myo}$ set to 0.01 and 0.001, respectively, empirically selected to balance the relative contributions of the cavity and myocardium. 

\subsection{Hyperparameters and training workflow}
\subsubsection{Training/testing split}
The 604 labeled 3D images representing data at ED and ES were divided into 504 paired images and labels for training and 100 paired images and labels for testing. For all splits, images from the same cycle (i.e., ED and ES frames) and acquisitions belonging to the same participant (i.e., scan and rescan) were grouped together.

\subsubsection{Stage I network training}
For the Stage I network, the 504 labeled images were further divided into a 70/30 split for training and validation, resulting in $n_{train}$ = 352 (176 cardiac cycles) and $n_{val}$ = 152 (76 cardiac cycles), respectively. The network was trained using an AdamW optimiser~\cite{Loshchilov2017DecoupledWD} with a learning rate of $1e^{-4}$ for 350 epochs, and a batch size of 4. The voxel encoder comprises increasing channels for each convolution block: 16, 32, 64, 128, with the bottleneck block consisting of 256 channels. The voxel decoder mirrors the encoder with a reversed number of channels. The bottleneck convolution block serves as the initial sampling point for the graph decoder network, moving progressively on each level of the voxel decoder. The Graph Feature Block utilises 32 channels for the graph convolutions. During training, a 20\% probability of augmentation was applied to the images and labels, including brightness and contrast adjustments, translation, rotation, uniform and non-uniform rescaling. To ensure the robustness of features sampled at the locations of template vertices, a small amount of random noise (modelled as a normal distribution with a standard deviation of 0.06, corresponding to $\approx 3\%$ translation) was applied to a random subset (20\%) of points in the initial template during feature sampling. Training required approximately 3.5 minutes per epoch on a Tesla~V100 GPU (32\,GB memory), and inference took about 0.09~seconds per image.

\subsubsection{Generation of pseudo ground truth meshes}
As ground truth meshes were only available at ED and ES frames, the Stage I network was used to generate meshes for the transitional (non-ED/ES) frames, which were subsequently treated as pseudo ground truth data for training the Stage II network. To generate reliable mesh predictions for all cases, an independent overfitted Stage I network was trained on the entire dataset (including ED and ES frames, for scan and rescan acquisitions), and subsequently used to predict meshes for all cases across all time frames. Each cardiac cycle ranged from 13 to 97 frames (mean ± SD of 38 ± 13 frames), resulting in a total of 11609 frames across 302 cardiac cycles. Importantly, the overfitted network was only used to generate the pseudo ground truth meshes, and was not used for any other purpose (such as generating intermediate features for the Stage II temporal network). 

\subsubsection{Data preparation for temporal network}
For the Stage II network, the input was designed to consist of the intermediate features obtained from the Stage I network. This approach reduces computational burden while allowing rich feature information to be extracted and processed. Features were extracted from layer \emph{l}=2 of the graph decoder block, resulting in a feature vector comprising of [$z_1$,~$x_2$,~$v_1$] for each vertex in each time frame. These features were extracted for all time frames and stacked together along the time dimension $T$, resulting in sequences $x_{feat}~\in~\mathbb{R}^{TxNxD}$, where $D$ corresponds to the total length of $z_1$, $x_2$, and $v_1$.

\subsubsection{Stage II network training}
To ensure optimal learning of the Stage II network from a diverse set of data without compounding errors, we employed the following approach to randomly split and mix the data. Of the 176 cardiac cycles in the Stage I training set, 140 were randomly assigned to training and 36 to validation. Similarly, of the 76 cycles in the Stage I validation set, 52 were assigned to training and 24 to validation. With this split, the training and validation sets for the Stage II network comprised of 192 and 60 cardiac cycles, respectively. Additionally, 50 cardiac cycles (corresponding with the 100 images in the Stage I test set) were held out for testing. The network was trained using an Adam optimiser with a learning rate of $1e^{-5}$ for 30000 epochs. A batch size of 16 was used, and the maximum number of frames ($T$) in the sequence was set to 64. To facilitate stable training, a learning rate warm-up strategy was employed for the first 2000 epochs, followed by a cosine learning rate scheduler. The Stage II temporal network was optimised for efficiency by design, which processes graphs as input and generates embeddings. The average training time per epoch was approximately 0.05 seconds, while the inference time for each image sequence was approximately 0.004 seconds.

\subsection{Validation and performance}
Model performance was evaluated on the held-out test set (n=50 cardiac cycles from 25 subjects). Evaluation metrics at ED and ES included the Dice coefficient for segmentation accuracy, and mean point-to-point (P2P) error (i.e., Euclidean distance between corresponding points), average surface distance (ASD), and Hausdorff Distance (HD) for mesh accuracy. To account for potentially increased P2P error due to rotational symmetry of the LV, we applied rigid Procrustes alignment between predicted and ground truth meshes to better understand the error contribution owing to differences in shape alone. Furthermore, we evaluated the trajectory smoothness of the predicted mesh vertices from both Stage I and II networks, as well as the pseudo ground truth vertices. Per-vertex trajectory smoothness was quantified using the second derivative of displacement across the time frames as an approximation of jitter~\cite{10.1007/978-3-030-87231-1_55}. The overall trajectory smoothness was assessed by averaging across all 1570 vertices.

The agreement in clinical cardiac indices was assessed based on GHOST-CAT model predictions and reference measurements from the paired CMR acquisitions. These included LV end-diastolic volume (EDV), end-systolic volume (ESV), LV mass (calculated as the average of mass at ED and ES, by multiplying the myocardial volume by a density of 1.055\,g/ml), ejection fraction (EF), and peak global longitudinal strain (GLS). Two-tailed paired \textit{t}-tests were conducted with significance defined as \emph{p}-values$<$0.05.

Finally, comparisons were performed against the following benchmark networks: nnU-Net (for segmentation), Voxel2Mesh (for both segmentation and mesh prediction), and MeshDeformNet (for mesh prediction). All benchmark methods were trained using the same training splits used for developing GHOST-CAT. The nnU-Net model was trained for one fold only for 300 epochs. Voxel2Mesh was trained to predict a single label (i.e., LV myocardium), and MeshDeformNet was trained to produce meshes of the myocardium and cavity. All benchmark models were trained using default configuration parameters as provided on their respective github repositories~\cite{Isensee2021NnU-Net:Segmentation, 10.1007/978-3-030-59719-1_30, KONG2021102222}. The results were evaluated quantitatively in terms of HD, ASD, and Dice coefficient on the LV myocardium. As there was no distinction between endocardial and epicardial surfaces, evaluations were performed on the continuous myocardial surface against the ground truth ED and ES meshes with 1570 vertices.

\section{Results}

\subsection{Qualitative assessment}
\begin{figure}[!t]
\centering
\includegraphics[width=\linewidth]{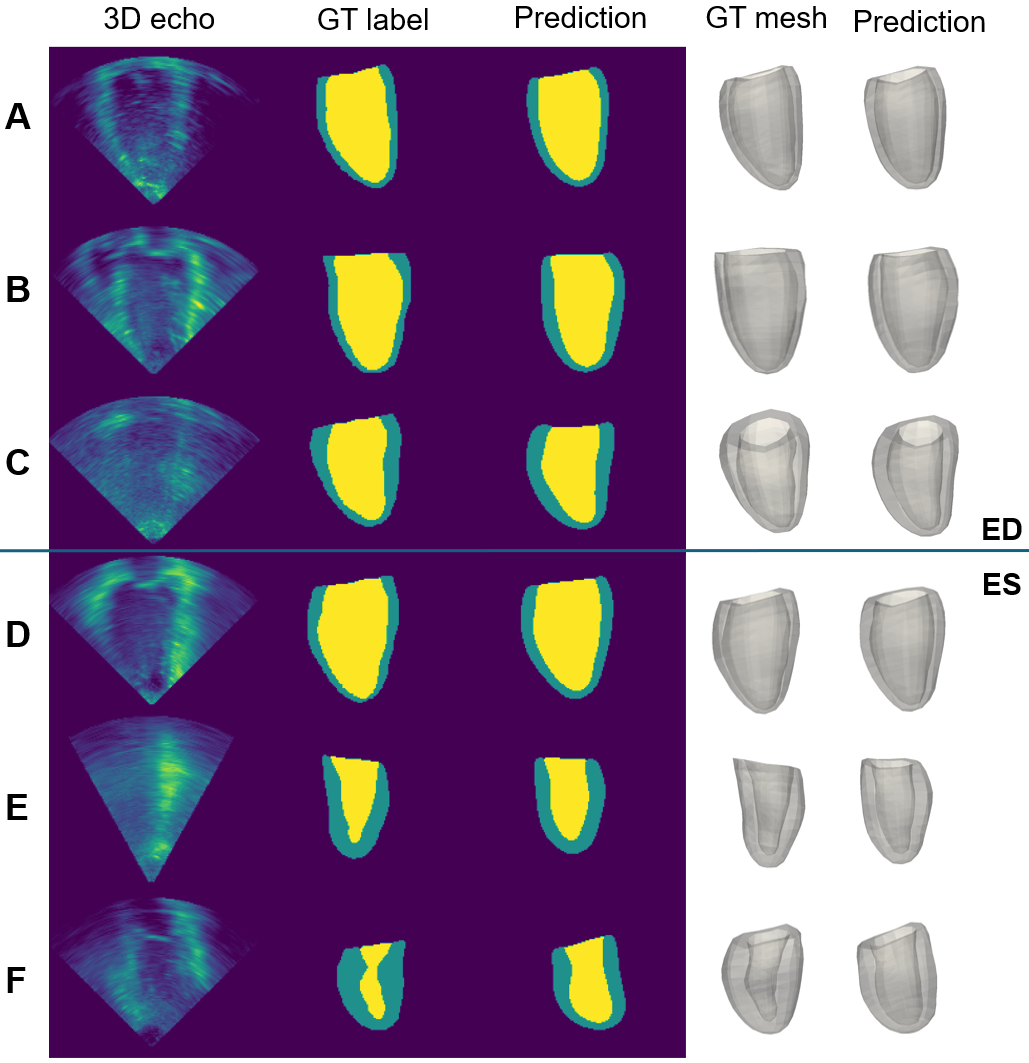}
\caption{Comparison of left ventricular (LV) segmentations and surface meshes by the GHOST-CAT Stage I network against ground truth (GT) labels for the (A) best, (B) median, and (C) worst test images at end-diastole (ED); and (D) best, (E) median, and (F) worst test images at end-systole (ES), selected based on the Dice coefficients of the LV cavity, which were assessed separately for ED and ES. The Dice coefficients for the myocardium and cavity at ED (rows A-C) were: 0.79 and 0.94 for the best case, 0.79 and 0.90 for the median case, and 0.68 and 0.85 for the worst case, respectively. For ES (rows D-F), the Dice coefficients for the myocardium and cavity were: 0.84 and 0.93 for the best case, 0.77 and 0.84 for the median case, and 0.77 and 0.65 for the worst case, respectively.}
\label{fig3}
\end{figure}

Fig.~\ref{fig3} presents qualitative comparisons between Stage I network predictions, and ground truth segmentations and meshes at ED and ES for the best, median, and worst test cases. From visual assessment, GHOST-CAT produced reasonable myocardium and cavity segmentations, as well as corresponding meshes of the LV myocardium.

\subsection{Segmentation and mesh accuracy}
Dice coefficients of segmentations on the test dataset (n=100 images) for GHOST-CAT Stage I, nnU-Net~\cite{Isensee2021NnU-Net:Segmentation}, and Voxel2Mesh~\cite{10.1007/978-3-030-59719-1_30} are presented in Table~\ref{tab1}.

\begin{table}[ht]
\centering
\caption{Dice coefficients for GHOST-CAT Stage I segmentations on the test dataset (n=100 images), in comparison with nnU-Net~\cite{Isensee2021NnU-Net:Segmentation} and Voxel2Mesh~\cite{10.1007/978-3-030-59719-1_30} benchmarks.}
\resizebox{\columnwidth}{!}{
\setlength{\tabcolsep}{5pt} 
\begin{tabular}{|l|c|c|c|}
\hline
Test (n=100) & nnU-Net & Voxel2Mesh & Stage I \\
\hline
Myocardium & 0.76 ± 0.06 & 0.71 ± 0.07 & 0.75 ± 0.07 \\
LV Cavity & 0.87 ± 0.05 & -- & 0.87 ± 0.05 \\
\hline
\end{tabular}%
}
\label{tab1}
\end{table}

Quantitative results of the mesh predictions by GHOST-CAT Stage I and II networks on the test dataset are reported in Table~\ref{tab2}. Rigid alignment led to significant improvement across all metrics for both the endocardium and epicardium for Stage~I and Stage~II networks (all \emph{p}-values$<$0.0001), suggesting that a substantial component of the distance errors for the non-aligned predictions is attributable to registration rather than shape. Notably, after alignment, the Stage~II network exhibited significantly lower mean errors than Stage~I across all metrics for both mesh surfaces.

\begin{table}[ht]
\centering
\caption{Evaluation of meshes on the test dataset (n=100 images), reported as point-to-point (P2P) error, Hausdorff distance (HD), and average surface distance (ASD) for GHOST-CAT Stage~I and Stage~II predictions, before and after rigid alignment. Asterisks (*) denote where Stage~II errors were statistically different from Stage~I for the same surface, metric, and alignment status. Values are mean $\pm$ standard deviation (mm).}
\resizebox{\columnwidth}{!}{%
\setlength{\tabcolsep}{5pt} 
\begin{tabular}{|l|l|c|c|c|}
\hline
\multicolumn{5}{|c|}{Before rigid alignment} \\
\hline
 & Surface & P2P & HD & ASD \\
\hline
Stage I  & Endocardium & 7.2 $\pm$ 1.9 & 9.4 $\pm$ 2.4 & 3.0 $\pm$ 0.6 \\
         & Epicardium  & 8.7 $\pm$ 2.4 & 10.2 $\pm$ 2.6 & 3.3 $\pm$ 0.5 \\
\hline
Stage II & Endocardium & 7.2 $\pm$ 1.7 & 9.6 $\pm$ 2.1 & 3.3 $\pm$ 0.6* \\
         & Epicardium  & 8.3 $\pm$ 2.1* & 10.1 $\pm$ 2.5 & 3.5 $\pm$ 0.5* \\
\hline
\multicolumn{5}{|c|}{After rigid alignment} \\
\hline
 & Surface & P2P & HD & ASD \\
\hline
Stage I  & Endocardium & 3.5 $\pm$ 0.9 & 7.6 $\pm$ 1.8 & 2.8 $\pm$ 0.6 \\
         & Epicardium  & 3.6 $\pm$ 0.8 & 7.7 $\pm$ 1.7 & 2.9 $\pm$ 0.5 \\
\hline
Stage II & Endocardium & 3.1 $\pm$ 0.8* & 7.2 $\pm$ 2.0* & 2.6 $\pm$ 0.6* \\
         & Epicardium  & 2.9 $\pm$ 0.8* & 7.3 $\pm$ 2.0* & 2.6 $\pm$ 0.6* \\
\hline
\end{tabular}%
}
\label{tab2}
\end{table}

\subsection{Comparison of clinical indices}
Table~\ref{tab3} shows the comparison of clinical indices derived from network predictions against reference CMR measurements. Across the three networks, the biases were less than 10\% of the mean CMR values for all measurements, with GHOST-CAT Stage II providing the lowest mean absolute differences in three out of five indices (ESV, LV mass, and GLS). Small but statistically significant biases in ejection fraction were observed for all three networks. 

\begin{table}
\centering
\caption{Evaluation of clinical indices for the test dataset (n=50 cardiac cycles), including end-diastolic volume (EDV), end-systolic volume (ESV), left ventricular (LV) mass, ejection fraction (EF), and global longitudinal strain (GLS). Asterisks (*) denote statistically significant differences from reference cardiac magnetic resonance (CMR) measurements. Values in bold indicate the smallest mean absolute difference for each index.}
\resizebox{\columnwidth}{!}{%
\setlength{\tabcolsep}{5pt}
\begin{tabular}{|l|c|c|c|c|}
\hline
n = 50 & CMR & nnU-Net & Stage I & Stage II \\
\hline
EDV (ml) & 162 ± 56  & -3 ± 13  & \textbf{-2 ± 16} & -11 ± 14* \\
ESV (ml) &  70 ± 48 & 3 ± 10  & 5 ± 12* & \textbf{-1 ± 13} \\
EF (\%) &  60 ± 11 & \textbf{-3 ± 5*}  & -4 ± 6* & -3 ± 6* \\
LV mass (g) &  131 ± 51 & 6 ± 12* &  7 ± 16* & \textbf{2 ± 19} \\
GLS (\%) & -18.9 ± 4.2 & N/A & 0.3 ± 2.9 & \textbf{0.2 ± 2.9} \\
\hline
\end{tabular}
}
\label{tab3}
\end{table}

\subsection{Temporal coherence}
\begin{figure*}[!t]
\centering
\includegraphics[width=130mm]{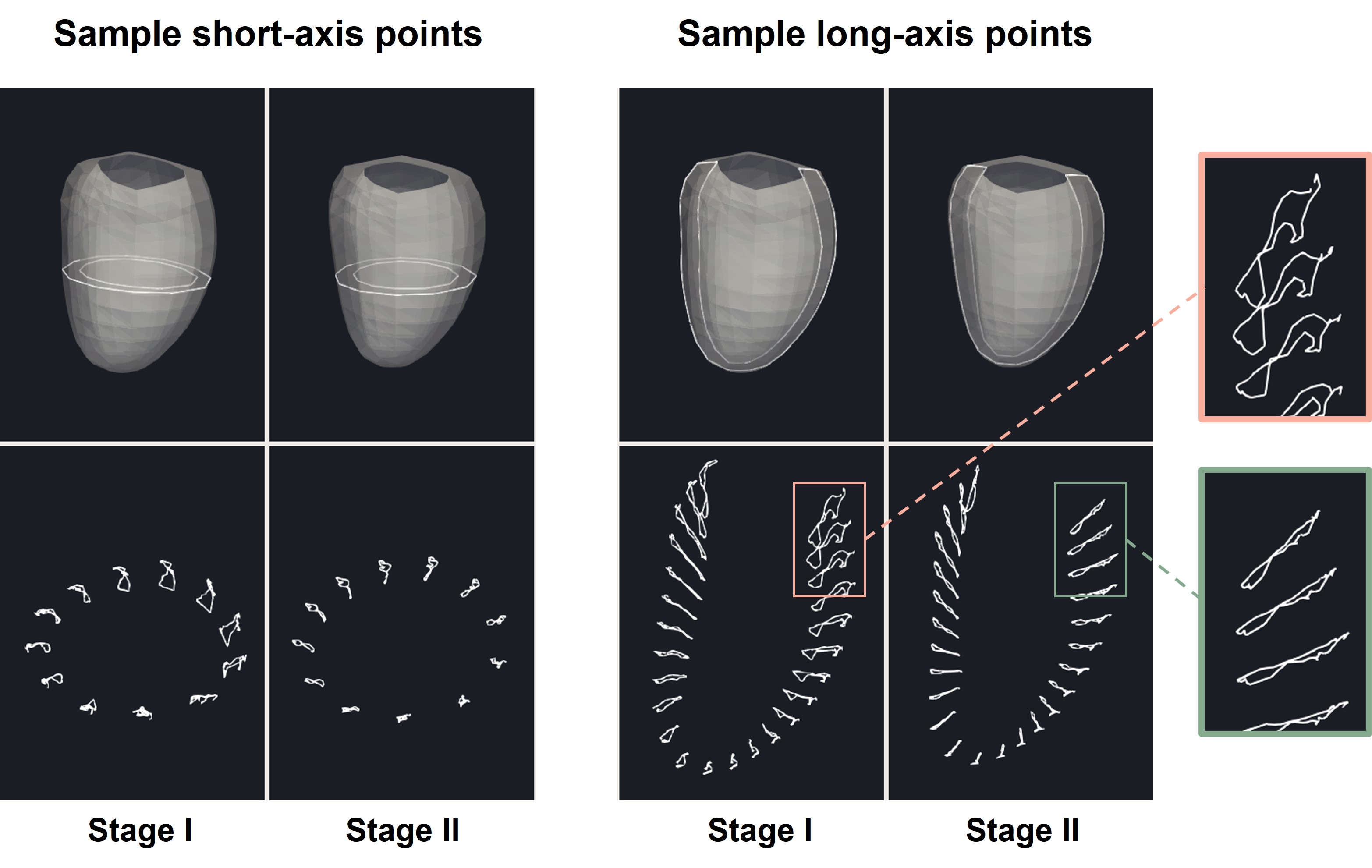}
\caption{Comparison of vertex trajectories (white lines) of the median test case between Stage I and Stage II networks. Lower panels show subsets of vertices sampled from the epicardial surface for an equatorial short-axis slice (left), and from the endocardial surface for a mid-ventricular longitudinal slice (right). Trajectories for the Stage II prediction (lower inset) were smoother than those for the Stage I prediction (upper inset).
 }
\label{fig4}
\end{figure*}

Smoother LV motion over the cardiac cycle was observed after applying the Stage II network (see Supplementary Video). Values for trajectory jitter across the test cases were 1.4 ± 0.5, 1.6 ± 0.6, and 1.0 ± 0.4, for pseudo ground truth, Stage I, and Stage II vertices, respectively. The differences in trajectory jitter for all pairwise comparisons were statistically significant. An illustrative example of vertex trajectories given in Fig.~\ref{fig4} demonstrates the reduction in apparent jitter associated with the Stage II network.

\begin{figure*}[!t]
\centerline{\includegraphics[width=140mm]{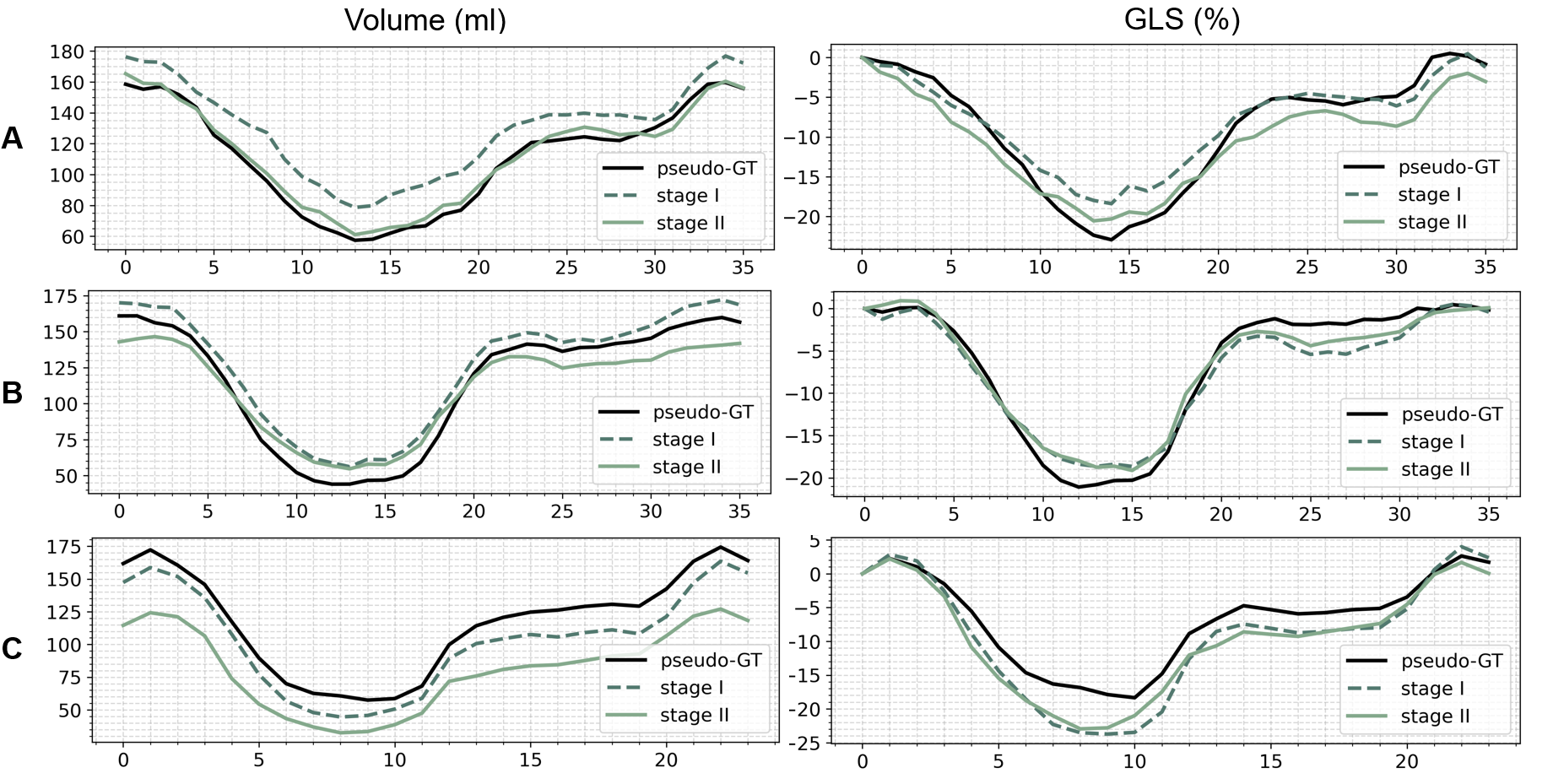}}
\caption{Left ventricular volume and global longitudinal strain (GLS) curves for the best (A), median (B), and worst (C) test cases. The plots show the comparison between results from the pseudo ground truth (GT), Stage I network, and after applying Stage II network.}
\label{fig5}
\end{figure*}

To assess the impact of the apparent improvement in temporal coherence in volume and GLS measurements, volume and strain curves were evaluated for the test cases. The resultant curves corresponding to the best, median, and worst cases (based on the average absolute volume differences between the Stage II and pseudo ground truth meshes across all time frames) are shown in Fig.~\ref{fig5}. Overall, both Stage I and Stage II networks produced curves that captured the dynamic features of systolic contraction and diastolic relaxation.

\subsection{Benchmark comparisons and mesh quality}
\begin{figure*}[!t]
\centering
\includegraphics[width=130mm]{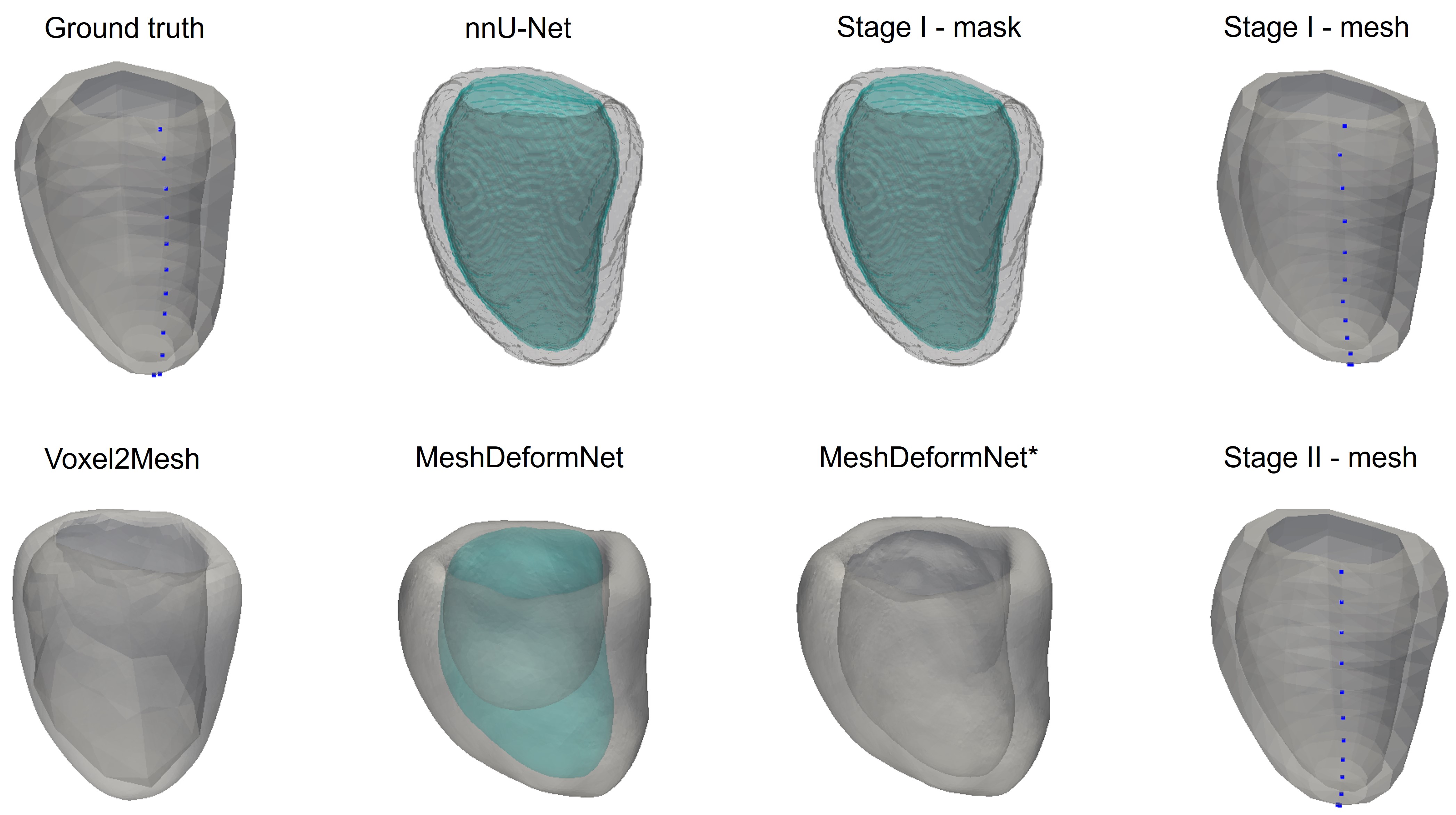}
\caption{Comparison of meshes generated for the median test case by the different methods. nnU-Net and GHOST-CAT Stage I produce segmentations of the left ventricular (LV) myocardium and cavity (teal shading). MeshDeformNet generates separate meshes for each structure, which intersect. A corrected version (marked with *) was obtained by removing the intersecting cavity region from the myocardium. Corresponding points on ground truth and GHOST-CAT meshes are indicated by dark blue markers.}
\label{fig6}
\end{figure*}

Meshes generated by GHOST-CAT were compared to existing benchmark methods to assess differences in performance. A visual comparison is presented in Fig.~\ref{fig6}. Of note, all meshes generated by MeshDeformNet initially exhibited intersections between the LV myocardium and cavity. To address this issue, intersecting regions were removed prior to evaluation to enable a more consistent comparison across methods. Table~\ref{tab4} presents corresponding quantitative metrics across the test dataset in terms of surface distance errors and Dice coefficients relative to the ground truth meshes derived from CMR, evaluated across the whole domain of the myocardium (i.e., without differentiation between endocardial and epicardial surfaces). As shown, GHOST-CAT Stage~I achieved the best average performance across all metrics, with Stage~II following closely behind (differing by only 0.1~mm in mean ASD and HD). Although Voxel2Mesh yielded a marginally higher mean Dice coefficient than the GHOST-CAT Stage II network, the resultant endocardial geometry was often unrealistic (as illustrated in Fig.~\ref{fig6}), likely leading to the higher HD values.

\begin{table}[t]
\caption{Quantitative evaluation of mesh reconstructions on the test dataset (n=100). Each method was evaluated against the ground truth meshes (1570 vertices per mesh). Metrics were computed over the entire myocardial surface (endocardium and epicardium) for each case.}
\centering
\setlength{\tabcolsep}{5pt}
\begin{tabular}{|l|c|c|c|}
\hline
n = 100 & ASD (mm) & HD (mm) & Dice \\
\hline
GHOST-CAT Stage I & 3.0 ± 0.5 & 9.9 ± 2.3 & 0.74 ± 0.07 \\
GHOST-CAT Stage II & 3.1 ± 0.4 & 10.0 ± 2.2 & 0.69 ± 0.10 \\
Voxel2Mesh & 3.1 ± 0.5 & 11.6 ± 2.2 & 0.71 ± 0.07 \\
MeshDeformNet & 7.4 ± 1.5 & 22.2 ± 4.9 & 0.41 ± 0.22 \\
MeshDeformNet (corrected) & 7.0 ± 1.2 & 21.6 ± 5.1 & 0.38 ± 0.24 \\
\hline
\end{tabular}
\label{tab4}
\end{table}

\section{Discussion}

Mesh reconstruction of the LV myocardium is an important task in cardiac image analysis, enabling accurate anatomical visualisation and quantitative assessment of cardiac function. In this study, we present GHOST-CAT, a two-stage deep learning framework that leverages CNN, GCN, and transformer architectures to reconstruct temporally coherent LV surface meshes directly from 3D echocardiograms. Evaluation results indicate that GHOST-CAT produces accurate and robust mesh reconstructions despite the challenges of noise, acoustic dropout, and limited image contrast, outperforming other state-of-the-art methods. This provides a scalable foundation for advanced modelling and potential clinical applications.

\subsection{Methodological insights}
We employed a hybrid approach combining convolutional and graph neural networks, using a predefined template with fixed topology. A second-stage network was then applied to refine the predictions in the temporal domain, which helped to ensure that reconstructed meshes maintained point correspondence with the template while remaining temporally coherent across the cardiac cycle.

In the Stage I network, we modified the layers to produce zero-centered extracted feature values. Although this change may seem minor, it facilitated faster training of the Stage II network and helped to reduce overfitting. Additionally, we removed the use of LNS as the feature sampling strategy, as we observed that the predicted meshes tended to converge toward an average shape. We hypothesise that this behaviour resulted from maintaining a constant number of vertices at each level without intermediate subdivisions, which may not be optimal for LNS. Therefore, we opted for a simpler feature sampling strategy using trilinear interpolation, which yielded more stable and accurate results in our experiments.

As in Voxel2Mesh~\cite{10.1007/978-3-030-59719-1_30} and MeshDeformNet~\cite{KONG2021102222}, we adopted a progressive mesh deformation approach. While Voxel2Mesh relies on Laplacian regularisation to control deformations, our regularisation constrained the amount of displacement at each decoder step. As we move up the image decoder module, the voxel feature maps progressively increase in spatial resolution while decreasing in the level of abstraction. In this setup, high-level abstractions primarily guide global mesh structure and large deformations, while low-level abstractions (i.e., higher resolution feature maps) contribute to smaller deformation refinements. We observed that high-level features were effective in deforming the template mesh toward approximately correct positions, with subsequent blocks providing incremental refinements. However, the final deformation block contributed little to no additional improvement, likely due to insufficient spatial features sampled from the high-resolution grid. This suggests that while feature sampling and GCNs performs reasonably well for spatial problems, better integration of spatial features into the graph network is necessary, and may lead to further improvements.

In contrast to other studies that employ generic sphere-like templates, we opted for a template mesh that more closely resembles the anatomical shape of the LV. This choice was made to facilitate faster convergence during training. Additionally, we experimented with an alternative template derived from the vertex-wise average of all ground-truth meshes. The network trained with this averaged template achieved comparable reconstruction accuracy. However, we observed that the learned weights were highly dependent on the specific template used during training, and did not generalise well when a different template was substituted at inference. In an attempt to mitigate overfitting of features sampled at the initial vertex locations, random noise was added to vertex positions during training to prevent consistent extraction of features at identical locations in the bottleneck layer. Despite this approach, the use of a different template shape (even in keeping the same structure and number of vertices) during inference resulted in poor performance, necessitating retraining. While the GCN component works relatively well for template deformation, its ability to generalise across spatial configurations remains limited. Such behaviour appears to be an inherent weakness of graph networks, even when explicit spatial features and coordinates are provided as input to each GCN layer. The persistent template-dependency remains a limiting factor, and additional work is needed to explore strategies for improving template invariance in mesh reconstruction frameworks. 

Our Stage II network operates efficiently by utilising graph-based features as inputs, as opposed to high-dimensional 3D image sequences. Although the network was trained with more than 10,000 images, these correspond to only about 250 unique temporal sequences, leading to a limited number of independent training samples and subsequent signs of overfitting. A more effective training strategy and additional data augmentation are needed to improve generalisation. However, in the current setup, end-to-end training is not feasible as it would incur impractical computational costs.

For the transformer, only MSE was used for optimisation, while the remaining mesh loss terms employed in Stage I were not included. Interestingly, the network was able to produce temporally consistent meshes even when the pseudo ground truth data lacked temporal coherence. This likely arises from the transformer’s inherent ability to process multiple features simultaneously through its attention mechanism. A volume loss term was introduced to penalise deviation of the predicted cavity size from that of the pseudo ground truth (even when the meshes were temporally smoothed). In addition, a regularisation term was added to promote mass conservation across frames. While these temporal constraints promote smooth and coherent motion, they may also introduce minor geometric distortions (possibly reflected by the lower Dice coefficients in Table~\ref{tab4}). 

The robustness of the Stage II network is further demonstrated by its ability to reconstruct temporally coherent meshes from graph features alone. Despite never having been exposed to any dense spatial representations (e.g., uniform grid-like features), the network was nevertheless able to infer spatially coherent meshes. This provides further evidence that the features extracted from the Stage I network contain sufficient information for accurate mesh reconstruction, and that using extracted features as the input to the Stage II network is a viable and effective strategy. Of note, the selection of features from layer \emph{l}=2 of the Graph Decoder Block was empirical based on the observation that the features from this layer resulted in substantial deformation towards the target shape, while features from subsequent layers appeared to primarily serve to refine vertex positions.

\subsection{Comparison with benchmarks}
Compared with the benchmark segmentation method (nnU-Net) used in MITEA~\cite{Zhao2023MITEA:Imaging}, GHOST-CAT achieves comparable segmentation performance. Moreover, our approach is preferable to Voxel2Mesh, as it produces surface meshes with point correspondence. The GHOST-CAT Stage I network efficiently handles both segmentation and mesh reconstruction tasks, while the Stage II network promotes temporal consistency, which is important for downstream applications such as strain analysis. The Stage I network achieved a 3\% improvement in Dice coefficient compared to Voxel2Mesh, while employing a similar architecture. In terms of mesh reconstruction, Voxel2Mesh and MeshDeformNet typically generated less accurate predictions, particularly with respect to the endocardial surface. While MeshDeformNet can produce multiple high-resolution meshes simultaneously, these often intersect, producing anatomically implausible myocardial geometries. 

Additionally, we investigated feature sampling from the voxel encoder, similar to that performed for MeshDeformNet~\cite{KONG2021102222}, while keeping GHOST-CAT Stage I network architecture unchanged. This resulted in a 1\% decrease in Dice coefficient for both the LV cavity and myocardium, and a 0.2 mm increase in ASD for both endocardial and epicardial vertices. Although these differences are likely acceptable in scenarios prioritising reduced computation, sampling from the decoder remains preferable as it incorporates deeper layer features and skip connections, leading to better overall performance.

\subsection{Limitations and future work}
While these results are compelling, a number of limitations should be acknowledged. Firstly, our method relies on ground truth meshes with point correspondence. Unlike Voxel2Mesh, which only requires segmentation labels (from which meshes can be derived directly), our framework necessitates a predefined template that restricts flexibility. Nevertheless, topological consistency is desirable for several downstream applications, such as statistical shape modelling. 

In terms of the template mesh, there are currently only 785 vertices per surface, which may be insufficient to adequately represent cases with complex myocardial anatomies. Increasing vertex density in both the template and ground truth meshes could improve reconstruction fidelity with only minor network modifications. Additionally, vertex positions in the template correspond to approximate anatomical landmarks, which limits the possible augmentation strategies and ranges. For instance, aggressive transformations (e.g., rotations greater than 30°) may not reflect the variation seen in real data due to physical image acquisition constraints and are therefore unrealistic. Ideally, the network should instead learn to infer vertex positions from image features directly, allowing for greater robustness to orientation and positional variability. Achieving this capability will require further investigation into landmark-independent feature learning and equivariant network designs.

A further limitation is the reliance on pseudo ground truth meshes generated by an overfitted GHOST-CAT Stage I network. As this network was trained using only the ED and ES frames, predictions for intermediate frames may include interpolation errors or unrealistic deformations that would be propagated to the Stage II network. Obtaining fully annotated sequences or incorporating self-supervised learning strategies may help to capture temporal dynamics more accurately. Likewise, expanding the training dataset to include a broader range of anatomical variations and pathological cases would further strengthen model robustness and clinical applicability.

Future work should focus on improving template independence through advanced augmentation, regularisation, and adaptive vertex initialisation strategies to enhance generalisation. Furthermore, integrating spatial features more effectively into the GCN, such as through 3D feature unfolding or attention-based fusion, may yield more accurate mesh reconstructions. Collectively, these developments are expected to enhance the overall performance and generalisability of GHOST-CAT.

\section*{Acknowledgment}
We acknowledge our clinical collaborators and the staff at the Centre for Advanced MRI (CAMRI), University of Auckland, for their expertise and assistance with the imaging components of this study. We also extend our gratitude to the participants who contributed their data to MITEA, without whom this research would not have been possible.

\bibliographystyle{IEEEtran}
\bibliography{references}

@article{KONG2021102222,
    title = {{A deep-learning approach for direct whole-heart mesh reconstruction}},
    year = {2021},
    journal = {Medical Image Analysis},
    author = {Kong, Fanwei and Wilson, Nathan and Shadden, Shawn},
    pages = {102222},
    volume = {74},
    doi = {10.1016/j.media.2021.102222},
    issn = {1361-8415}
}

@inproceedings{Loshchilov2017DecoupledWD,
    title = {{Decoupled Weight Decay Regularization}},
    year = {2017},
    booktitle = {International Conference on Learning Representations},
    author = {Loshchilov, Ilya and Hutter, Frank}
}

@article{10054471,
    title = {{Dual-Branch TransV-Net for 3-D Echocardiography Segmentation}},
    year = {2023},
    journal = {IEEE Transactions on Industrial Informatics},
    author = {Zhang, Jiapeng and Wang, Yongxiong and Chen, Lijun and Liu, Jinlong and Zhang, Sunjie and Pan, Zhiqun and Wang, Zhe and Tang, Zhenhui and Guo, Ying},
    number = {12},
    pages = {11675--11686},
    volume = {19},
    doi = {10.1109/TII.2023.3249904}
}

@article{GORCSAN20111401,
    title = {{Echocardiographic Assessment of Myocardial Strain}},
    year = {2011},
    journal = {Journal of the American College of Cardiology},
    author = {Gorcsan, John and Tanaka, Hidekazu},
    number = {14},
    pages = {1401--1413},
    volume = {58},
    doi = {10.1016/j.jacc.2011.06.038},
    issn = {0735-1097}
}

@inproceedings{10306804,
    title = {{Efficient Left Ventricle Segmentation in 3D Echocardiography using Deep nnU-Net}},
    year = {2023},
    booktitle = {2023 IEEE International Ultrasonics Symposium (IUS)},
    author = {Akbari.S, Somayeh and Papangelopoulou, Konstantina. and Munteanu-Mirea, Oana and Queir{\'{o}}s, Sandro and D’Hooge, Jan},
    pages = {1--4},
    doi = {10.1109/IUS51837.2023.10306804}
}

@inproceedings{10.1145/37401.37422,
    title = {{Marching cubes: A high resolution 3D surface construction algorithm}},
    year = {1987},
    booktitle = {Proceedings of the 14th Annual Conference on Computer Graphics and Interactive Techniques},
    author = {Lorensen, William E and Cline, Harvey E},
    pages = {163–169},
    series = {SIGGRAPH '87},
    publisher = {Association for Computing Machinery},
    address = {New York, NY, USA},
    isbn = {0897912276},
    doi = {10.1145/37401.37422}
}

@article{Lei2023medlam,
    title = {{MedLSAM: Localize and Segment Anything Model for 3D Medical Images}},
    year = {2023},
    journal = {arXiv preprint arXiv:},
    author = {Wenhui Lei Xu Wei, Xiaofan Zhang Kang Li Shaoting Zhang}
}

@inproceedings{10.1007/978-3-031-43990-2_17,
    title = {{ModusGraph: Automated 3D and 4D Mesh Model Reconstruction from Cine CMR with Improved Accuracy and Efficiency}},
    year = {2023},
    booktitle = {Medical Image Computing and Computer Assisted Intervention -- MICCAI 2023},
    author = {Deng, Yu and Xu, Hao and Rodrigo, Sashya and Williams, Steven E and Williams, Michelle C and Niederer, Steven A and Pushparajah, Kuberan and Young, Alistair},
    editor = {Greenspan, Hayit and Madabhushi, Anant and Mousavi, Parvin and Salcudean, Septimiu and Duncan, James and Syeda-Mahmood, Tanveer and Taylor, Russell},
    pages = {173--183},
    publisher = {Springer Nature Switzerland},
    address = {Cham},
    isbn = {978-3-031-43990-2}
}

@article{DBLP:journals/corr/abs-1804-01654,
    title = {{Pixel2Mesh: Generating 3D Mesh Models from Single {\{}RGB{\}} Images}},
    year = {2018},
    journal = {CoRR},
    author = {Wang, Nanyang and Zhang, Yinda and Li, Zhuwen and Fu, Yanwei and Liu, Wei and Jiang, Yu-Gang},
    volume = {abs/1804.0},
    url = {http://arxiv.org/abs/1804.01654}
}

@misc{cheng2023sammed2d,
    title = {{SAM-Med2D}},
    year = {2023},
    author = {Cheng, Junlong and Ye, Jin and Deng, Zhongying and Chen, Jianpin and Li, Tianbin and Wang, Haoyu and Su, Yanzhou and Huang, Ziyan and Chen, Jilong and Sun, Lei Jiangand Hui and He, Junjun and Zhang, Shaoting and Zhu, Min and Qiao, Yu},
    arxivId = {cs.CV/2308.16184}
}

@misc{wang2023sammed3d,
    title = {{SAM-Med3D}},
    year = {2023},
    author = {Wang, Haoyu and Guo, Sizheng and Ye, Jin and Deng, Zhongying and Cheng, Junlong and Li, Tianbin and Chen, Jianpin and Su, Yanzhou and Huang, Ziyan and Shen, Yiqing and Fu, Bin and Zhang, Shaoting and He, Junjun and Qiao, Yu},
    arxivId = {cs.CV/2310.15161}
}

@article{kirillov2023segany,
    title = {{Segment Anything}},
    year = {2023},
    journal = {arXiv:2304.02643},
    author = {Kirillov, Alexander and Mintun, Eric and Ravi, Nikhila and Mao, Hanzi and Rolland, Chloe and Gustafson, Laura and Xiao, Tete and Whitehead, Spencer and Berg, Alexander C and Lo, Wan-Yen and Doll{\'{a}}r, Piotr and Girshick, Ross}
}

@article{MedSAM,
    title = {{Segment Anything in Medical Images}},
    year = {2024},
    journal = {Nature Communications},
    author = {Ma, Jun and He, Yuting and Li, Feifei and Han, Lin and You, Chenyu and Wang, Bo},
    pages = {1--9},
    volume = {15}
}

@inproceedings{10.1007/978-3-030-59719-1_30,
    title = {{Voxel2Mesh: 3D Mesh Model Generation from Volumetric Data}},
    year = {2020},
    booktitle = {Lecture Notes in Computer Science (including subseries Lecture Notes in Artificial Intelligence and Lecture Notes in Bioinformatics)},
    author = {Wickramasinghe, Udaranga and Remelli, Edoardo and Knott, Graham and Fua, Pascal},
    editor = {Martel, Anne L and Abolmaesumi, Purang and Stoyanov, Danail and Mateus, Diana and Zuluaga, Maria A and Zhou, S Kevin and Racoceanu, Daniel and Joskowicz, Leo},
    pages = {299--308},
    volume = {12264 LNCS},
    publisher = {Springer International Publishing},
    address = {Cham},
    isbn = {9783030597184},
    doi = {10.1007/978-3-030-59719-1{\_}30},
    issn = {16113349},
    arxivId = {1912.03681}
}

@inproceedings{Vaswani2017AttentionNeed,
    title = {{Attention is all you need}},
    year = {2017},
    booktitle = {Advances in Neural Information Processing Systems},
    author = {Vaswani, Ashish and Shazeer, Noam and Parmar, Niki and Uszkoreit, Jakob and Jones, Llion and Gomez, Aidan N. and Kaiser, Lukasz and Polosukhin, Illia},
    pages = {5999--6009},
    volume = {2017-Decem},
    issn = {10495258},
    arxivId = {1706.03762}
}

@article{Gilbert2019IndependentStudy,
    title = {{Independent Left Ventricular Morphometric Atlases Show Consistent Relationships with Cardiovascular Risk Factors: A UK Biobank Study}},
    year = {2019},
    journal = {Scientific Reports},
    author = {Gilbert, Kathleen and Bai, Wenjia and Mauger, Charlene and Medrano-Gracia, Pau and Suinesiaputra, Avan and Lee, Aaron M. and Sanghvi, Mihir M. and Aung, Nay and Piechnik, Stefan K. and Neubauer, Stefan and Petersen, Steffen E. and Rueckert, Daniel and Young, Alistair A.},
    doi = {10.1038/s41598-018-37916-6},
    issn = {20452322},
    pmid = {30718635}
}

@article{Wang2018LeftAnalysis,
    title = {{Left Ventricular Diastolic Myocardial Stiffness and End-Diastolic Myofibre Stress in Human Heart Failure Using Personalised Biomechanical Analysis}},
    year = {2018},
    journal = {Journal of Cardiovascular Translational Research},
    author = {Wang, Zhinuo J and Wang, Vicky Y and Bradley, Chris P and Nash, Martyn P and Young, Alistair A and Cao, J Jane},
    number = {4},
    pages = {346--356},
    volume = {11},
    doi = {10.1007/s12265-018-9816-y},
    issn = {1937-5395}
}

@article{Zhao2023MITEA:Imaging,
    title = {{MITEA: A dataset for machine learning segmentation of the left ventricle in 3D echocardiography using subject-specific labels from cardiac magnetic resonance imaging}},
    year = {2023},
    journal = {Frontiers in Cardiovascular Medicine},
    author = {Zhao, Debbie and Ferdian, Edward and Maso Talou, Gonzalo D and Quill, Gina M and Gilbert, Kathleen and Wang, Vicky Y and Babarenda Gamage, Thiranja P and Pedrosa, João and D’hooge, Jan and Sutton, Timothy M and Lowe, Boris S and Legget, Malcolm E and Ruygrok, Peter N and Doughty, Robert N and Camara, Oscar and Young, Alistair A and Nash, Martyn P},
    month = {1},
    volume = {9},
    isbn = {2297-055X},
    doi = {10.3389/fcvm.2022.1016703},
    issn = {2297-055X}
}

@article{Isensee2021NnU-Net:Segmentation,
    title = {{nnU-Net: a self-configuring method for deep learning-based biomedical image segmentation}},
    year = {2021},
    journal = {Nature Methods},
    author = {Isensee, Fabian and Jaeger, Paul F and Kohl, Simon A A and Petersen, Jens and Maier-Hein, Klaus H},
    number = {2},
    pages = {203--211},
    volume = {18},
    doi = {10.1038/s41592-020-01008-z},
    issn = {1548-7105}
}

@article{Kipf2017Semi-supervisedNetworks,
    title = {{Semi-supervised classification with graph convolutional networks}},
    year = {2017},
    journal = {5th International Conference on Learning Representations, ICLR 2017 - Conference Track Proceedings},
    author = {Kipf, Thomas N. and Welling, Max},
    arxivId = {1609.02907}
}

@InProceedings{10.1007/978-3-030-87231-1_55,
	author={Gonzales, Ricardo A. and Lamy, J{\'e}r{\^o}me and Seemann, Felicia and Heiberg, Einar and Onofrey, John A. and Peters, Dana C.},
	editor={de Bruijne, Marleen and Cattin, Philippe C. and Cotin, St{\'e}phane and Padoy, Nicolas and Speidel, Stefanie and Zheng, Yefeng and Essert, Caroline},
	title={TVnet: Automated Time-Resolved Tracking of the Tricuspid Valve Plane in MRI Long-Axis Cine Images with a Dual-Stage Deep Learning Pipeline},
	booktitle={Medical Image Computing and Computer Assisted Intervention -- MICCAI 2021},
	year={2021},
	publisher={Springer International Publishing},
	address={Cham},
	pages={567--576},
	isbn={978-3-030-87231-1}
}

@article{GAGGION2025103630,
    title = {Multi-view hybrid graph convolutional network for volume-to-mesh reconstruction in cardiovascular MRI},
    journal = {Medical Image Analysis},
    volume = {104},
    pages = {103630},
    year = {2025},
    issn = {1361-8415},
    doi = {10.1016/j.media.2025.103630},
    author = {Nicolás Gaggion and Benjamin A. Matheson and Yan Xia and Rodrigo Bonazzola and Nishant Ravikumar and Zeike A. Taylor and Diego H. Milone and Alejandro F. Frangi and Enzo Ferrante},
}

@article{CHEN2021102228,
    title = {Shape registration with learned deformations for 3D shape reconstruction from sparse and incomplete point clouds},
    journal = {Medical Image Analysis},
    volume = {74},
    pages = {102228},
    year = {2021},
    issn = {1361-8415},
    doi = {10.1016/j.media.2021.102228},
    author = {Xiang Chen and Nishant Ravikumar and Yan Xia and Rahman Attar and Andres Diaz-Pinto and Stefan K Piechnik and Stefan Neubauer and Steffen E Petersen and Alejandro F Frangi},
}

@article{AHN2023102711,
    title = {Co-attention spatial transformer network for unsupervised motion tracking and cardiac strain analysis in 3D echocardiography},
    journal = {Medical Image Analysis},
    volume = {84},
    pages = {102711},
    year = {2023},
    issn = {1361-8415},
    doi = {https://doi.org/10.1016/j.media.2022.102711},
    author = {Shawn S. Ahn and Kevinminh Ta and Stephanie L. Thorn and John A. Onofrey and Inga H. Melvinsdottir and Supum Lee and Jonathan Langdon and Albert J. Sinusas and James S. Duncan},
}

@InProceedings{10.1007/978-3-031-94559-5_24,
    author="Fan, Yiling
    and Roche, Ellen",
    editor="Chabiniok, Radom{\'i}r
    and Zou, Qing
    and Hussain, Tarique
    and Nguyen, Hoang H.
    and Zaha, Vlad G.
    and Gusseva, Maria",
    title="Transformer-Based Surrogate Modeling for Efficient Left Ventricular Digital Twin",
    booktitle="Functional Imaging and Modeling of the Heart",
    year="2025",
    publisher="Springer Nature Switzerland",
    address="Cham",
    pages="260--274",
}

@InProceedings{10.1007/978-3-032-05325-1_33,
    author="Qiao, Mengyun
    and Zheng, Jin
    and Zhang, Weitong
    and Ma, Qiang
    and Li, Liu
    and Kainz, Bernhard
    and O'Regan, Declan P.
    and Matthews, Paul M.
    and Niederer, Steven
    and Bai, Wenjia",
    editor="Gee, James C.
    and Alexander, Daniel C.
    and Hong, Jaesung
    and Iglesias, Juan Eugenio
    and Sudre, Carole H.
    and Venkataraman, Archana
    and Golland, Polina
    and Kim, Jong Hyo
    and Park, Jinah",
    title="Mesh4D: A Motion-Aware Multi-view Variational Autoencoder for 3D+t Mesh Reconstruction",
    booktitle="Medical Image Computing and Computer Assisted Intervention -- MICCAI 2025",
    year="2026",
    publisher="Springer Nature Switzerland",
    address="Cham",
    pages="343--353",
    isbn="978-3-032-05325-1"
}

@Article{Qiao2025,
    author={Qiao, Mengyun
    and McGurk, Kathryn A.
    and Wang, Shuo
    and Matthews, Paul M.
    and O'Regan, Declan P.
    and Bai, Wenjia},
    title={A personalized time-resolved 3D mesh generative model for unveiling normal heart dynamics},
    journal={Nature Machine Intelligence},
    year={2025},
    month={May},
    day={01},
    volume={7},
    number={5},
    pages={800-811},
    issn={2522-5839},
    doi={10.1038/s42256-025-01035-5},
}

@article{HeartFormer,
    author = {Ma, Zhengda and Banerjee, Abhirup},
    year = {2025},
    month = {11},
    pages = {},
    title = {HeartFormer: Semantic-Aware Dual-Structure Transformers for 3D Four-Chamber Cardiac Point Cloud Reconstruction},
    journal = {arXiv preprint arXiv:},
    doi = {10.48550/arXiv.2512.00264}
}

\end{document}